\documentclass{article}
\usepackage[english]{babel}
\usepackage{amsmath}
\usepackage{amssymb}
\usepackage{example}
\usepackage{multibbl}
\usepackage{theorem}
\usepackage{relsize}
\usepackage{graphicx}
\usepackage{subfigure}
\graphicspath{{figss/}}
\usepackage{color}
\usepackage{pstricks,pst-all,pstricks-add,pst-plot,pst-eucl}
\usepackage{enumerate}
\usepackage[retainorgcmds]{IEEEtrantools}
\usepackage[labelfont=bf,labelsep=period,justification=raggedright]{caption}
\usepackage{enumerate}
\usepackage{soul}

\usepackage{setspace}

\definecolor{brown}{rgb}{0.64,0.16,0.16}
\definecolor{ForestGreen}{rgb}{0.13,0.54,0.13}
\definecolor{purple}{rgb}{0.62,0.12,0.94}
\definecolor{DodgerBlue}{rgb}{0.11,0.56,0.98}
\definecolor{RoyalBlue}{rgb}{0.25,0.41,0.88}
\definecolor{B}{rgb}{0,0,1}
\definecolor{G}{rgb}{0,0.502,0}
\definecolor{R}{rgb}{1,0,0}
\definecolor{cooperative}{rgb}{0.78,0.86,0.94}
\definecolor{competitive}{rgb}{0.68,0.72,0.78}

\usepackage{framed}
\usepackage{lipsum}

\makeatletter
\renewcommand{\fnum@figure}{\small\textbf{\figurename~\thefigure}}
\makeatother

\makeatletter
\renewcommand\@biblabel[1]{}
\makeatother

\theorembodyfont{\rm}

\title{\textbf{Modulation and Robustness of Endogenous Neuronal Spiking}\\
}
\author{Guillaume Drion$^{1,2,\ast}$, Alessio Franci$^{3,4}$, Vincent Seutin$^{2}$ \& Rodolphe Sepulchre$^{1,4}$\\
\small{$^1$Department of Electrical Engineering and Computer Science and GIGA Research, University of Li\`ege, B-4000 Li\`ege, Belgium.}\\
\small{$^2$Laboratory of Pharmacology and GIGA Neurosciences, University of Li\`ege, B-4000 Li\`ege, Belgium.}\\
\small{$^3$INRIA Lille-Nord Europe, Orchestron project, 59650 Villeneuve d'Ascq, France.}\\
\small{$^4$Department of Engineering, University of Cambridge, Cambridge CB2 1PZ, UK.}\\
\small{$\ast$ E-mail: gdrion@ulg.ac.be}}
\date{}

\begin{document}
\setulcolor{} 
\setstcolor{red}  
\sethlcolor{white} 

\maketitle



\section*{Abstract}
Neuronal spiking exhibits an exquisite combination of modulation and robustness properties, rarely matched in artificial systems. We exploit the particular interconnection structure of conductance based models to investigate this remarkable property. We find that much of neuronal modulation and robustness can be explained by separating the total transmembrane current into three different components corresponding to the three time scales of neuronal bursting. Each equivalent current aggregates many ionic contributions into an equivalent voltage-dependent conductance, which defines a key modulation parameter. Plugging those equivalent feedback gains in a minimal abstract model recovers many experimental modulation scenarii as modulatory paths in elementary two-parameter charts. Likewise, robustness owes to the many possible physiological realizations of a same equivalent conductance, highlighting the role of equivalent conductances as prominent targets for neuromodulation and intrinsic homeostasis.


\section{Introduction}
The regulation mechanisms of neuronal firing patterns are complex because they involve a variety of ionic currents that modulate excitability across a broad span of time-scales and activation ranges. This complexity is the source of many puzzles in computational neuroscience because extracting from detailed computational models the modulation and robustness principles observed in experimental neurophysiology is challenging. 

We attempt to extract from this complexity two parameters that largely govern the modulation of neuronal spiking. The proposed methodology is directly inspired from Hodgkin and Huxley modeling principles (Hodgkin and Huxley, 1952a,b): (i) voltage variations across the membrane equal the total ionic current through the membrane, (ii) the total current is the net sum of many voltage dependent currents, that act as parallel feedback loops, (iii) much insight about the temporal resolution of firing patterns comes from separating the aggregate current contribution in different time scales. We separate the total ionic current into three components: the fast component is in the time scale of spike depolarization, the slow component is in the time scale of the spike termination (or intraburst frequency) whereas the ultraslow component is in the time scale of slow firing (or interburst frequency). Our two key parameters are the feedback gains of the slow and ultraslow feedback loops, respectively.

We showed in recent work that the sign of the slow equivalent feedback gain determines an important switch in neuronal excitability (Franci et al., 2013a). A negative slow feedback gain corresponds to the restorative repolarization in HH model. In contrast, a positive slow feedback gain induces bistability in the slow time scale. This memory element is a critical component of bursting (Franci et al., 2013b). Here we show that the slow feedback gain ($\bar{g}_{s}$) and the ultraslow feedback gain ($\bar{g}_{us}$) are the key modulators not only of the transition from slow tonic spiking to bursting but also of physiologically distinct bursters. Remarkably, an abstract model that accounts for local deviations from the singular situation where both the slow and ultraslow feedback gains vanish suffices to reproduce the temporal modulation of all the distinct firing patterns and to make physiological predictions in two-parameter charts.

Predictions in the abstract two-parameter charts translate directly to physiological predictions in a given conductance-based model. Maximal conductances of currents are modulators when their variation affects the feedback gains. In contrast, the robustness of a firing pattern to large variations of conductances is predicted when they do not alter the two equivalent gains. We use a detailed computational model of a stomatogastric ganglion (STG, Goldman et al., 2001) to translate predictions from our abstract model to physiologically plausible modulations of maximal conductances and we show a striking similarity between the abstract and physiological modulation paths. To further investigate the physiological relevance of the proposed approach, we revisit the experimental data of two prominent modulators: the short-term modulation regulated by the transient deinactivation of T- type calcium channels in e.g. thalamocortical and subthalamic nucleus neurons and the long term modulation of serotonin in AplysiaÕs R15 neurons.

\section{Results}

\subsection*{Equivalent conductances determine feedback gains in each timescale}
\begin{figure}[h!]
\centering
 \includegraphics[width=0.95\linewidth]{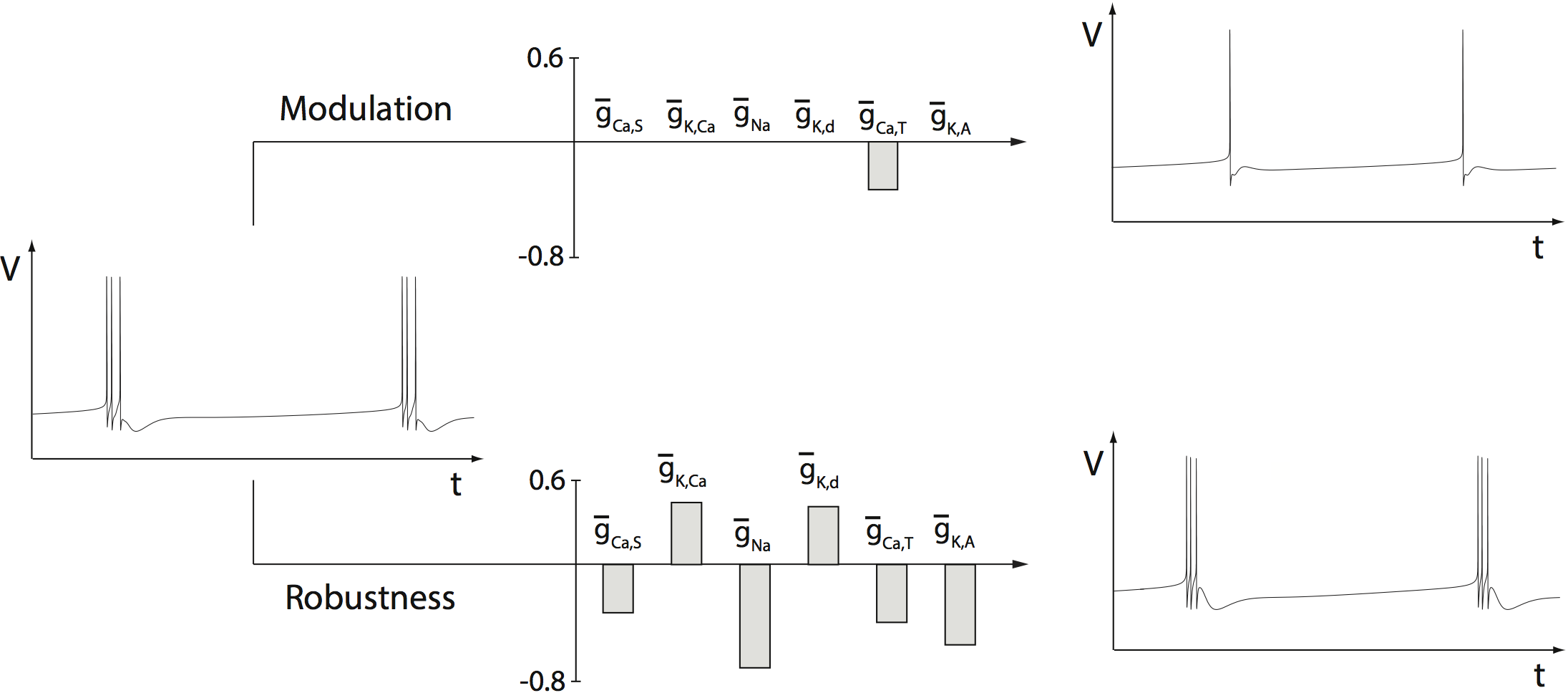}
\caption{\textbf{Sensitivity and robustness of neuronal spiking.} The picture shows membrane potential evolution over time of a STG neuron model for three different parameter sets (values represent the relative variations in each maximal conductance). The upper right trace shows a drastically different behavior than left trace although it only mildly differ in one parameter. On the other hand, left and lower right traces show very similar behaviors although the parameter sets strongly differ. These examples highlight concomitant sensitivity and robustness of neuron signaling to parameter heterogeneity.}\label{FIG:1}
\end{figure}

Figure \ref{FIG:1} illustrates an apparent paradox of neuronal rhythmicity: small variations of a single conductance suffice to alter the rhythm, suggesting a high sensitivity - or modulation capabilities -, whereas the concomitant variation of several conductances may leave the rhythm unchanged, suggesting low sensitivity - or robustness.

The combined modulation and robustness properties of ionic conductances owe to the particular parallel feedback structure of ionic currents determining the voltage variations of the plasma membrane: due to the diversity of ion channels, many different currents contribute, but only their net sum determines the voltage variation at a given instant of time. Because most of those currents are voltage-gated, they act as the parallel interconnection of nonlinear feedback amplifiers, where each conductance acts as a voltage-dependent feedback gain. The popular I/V curve in electrophysiology is an illustration of the resulting equivalent feedback loop in static conditions, which is not sufficient to capture the temporal mechanisms of spiking activity. This curve only extracts the static current $I_{static}$ generated by a persistent opening state of different channels at steady-state, but does not account for the dynamical properties of the neuron. 

Early in their investigations, Hodgkin and Huxley advised to separate the current contributions in different time-scales to resolve the temporal variations of electrical activity. Following this strategy, we separate the total current in four well separated temporal contributions: $I_{static}$, the persistent static current that sets the steady-state potential $V_{ss}$, modulated by all channels that are not completely closed at resting state ; $I_{fast}$ (or $I_{f}$), the analog of $I_{early}$ in the original work of Hodgkin and Huxley (Hodgkin and Huxley, 1952a), modulated by fast channels and accounting for the rapid upstroke of the action potential ; $I_{slow}$ (or $I_{s}$) ($I_{late}$ in Hodgkin and Huxley, 1952a), modulated by slow channels and accounting for the downstroke of action potential and the after-spike repolarization ; and $I_{ultraslow}$ (or $I_{us}$) (not considered in Hodgkin and Huxley, 1952a), modulated by ultraslow channels and accounting for the modulation of the neuronal rhythm over the course of many action potentials. Our main ansatz is that ionic currents can be combined within each time-scale and that separating the contribution of $I_{static}$, $I_f$, $I_s$ and $I_{us}$ suffices to understand the modulation and robustness of neuronal activity.

\begin{figure}[h!]
\centering
 \includegraphics[width=1\linewidth]{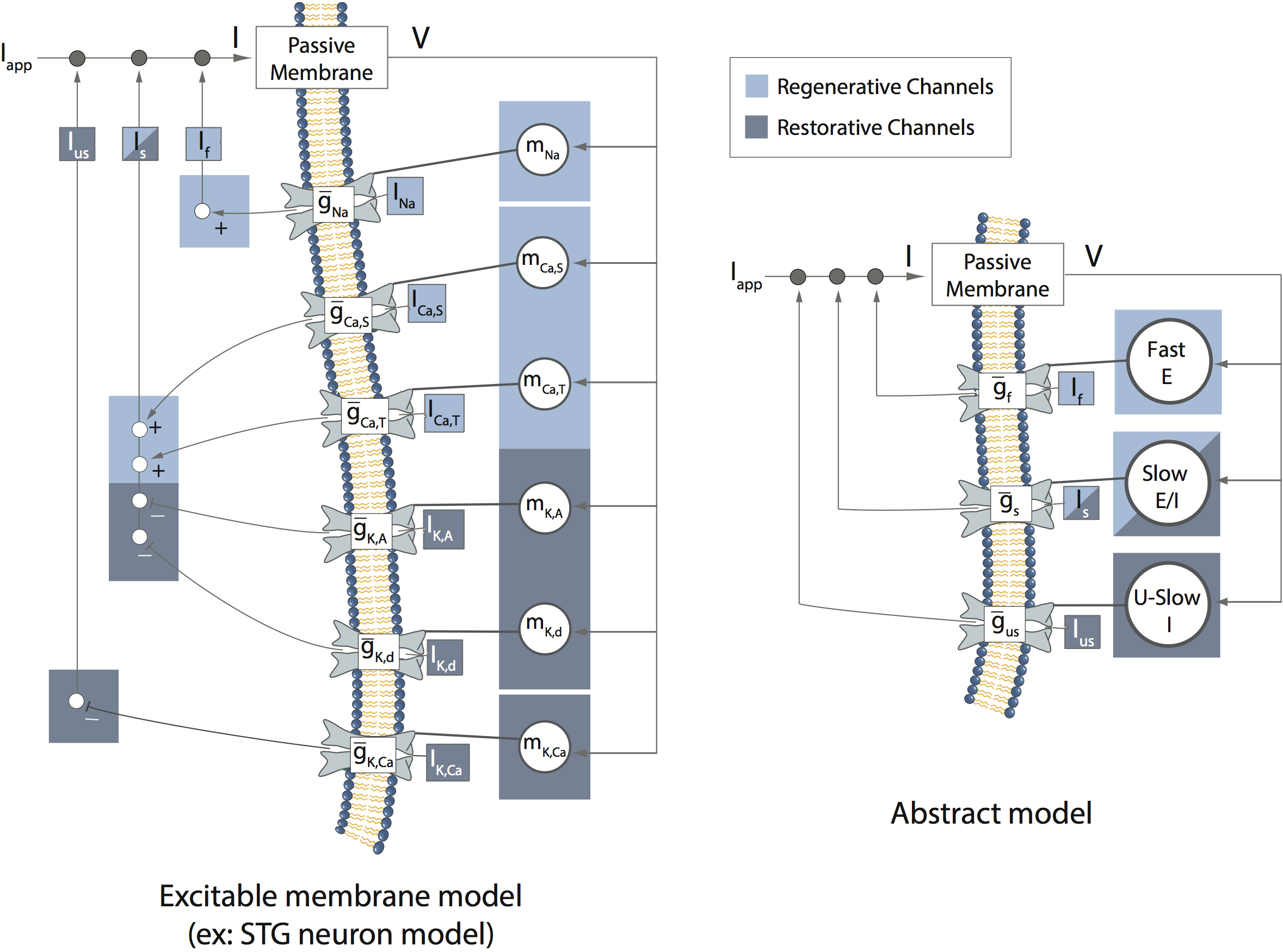}
\caption{\textbf{Time-scale classification of the feedback loops generated by the different ionic currents in a neuron model.} \textbf{Left}, classification of the contributions of the six ionic currents present in the STG neuron model into three feedback loops. \textbf{Right}, abstract equivalent model reduced to three equivalent gains (E=positive feedback, I=negative feedback, E/I=mixed feedback). }\label{FIG:2}
\end{figure}

Figure \ref{FIG:2}, left illustrates the time-scale classification of the feedback loops generated by six distinct ionic currents of a stomatogastric ganglion (STG) neuron model according to the activation kinetics. We chose to illustrate many of our results on this STG neuron model because it has served as a prototype model for many computational and experimental investigations on neuromodulation and homeostasis (Maclean et al., 2003; Schultz et al., 2006; Marder, 2011; Zhao et al., 2012). Figure \ref{FIG:2}, right sketches the abstract equivalent model reduced to three equivalent feedback gains. The fast feedback gain mainly accounts for the fast autocatalysis achieved by sodium activation. Spike generation relies on this positive feedback. Experiments that block transient (tetrodotoxin sensitive) sodium channels reduce this gain and often result in the absence of spikes. Because we focus on physiological mechanisms of neuronal spiking modulation, we fix the parameter $\bar{g}_f$ throughout the paper and choose it large enough to enable spike generation. 

The slow feedback gain accounts for the net sum of the many slow currents. Slow currents can either be restorative (i.e. providing a negative feedback to membrane potential variations) or regenerative (i.e. providing a positive feedback to membrane potential variations). Slow restorative channels (two potassium channels in the STG example) are essential for spike downstroke and repolarization period. But the presence of slow regenerative channels (two calcium channels in the STG example) is an essential source of neuromodulation because it makes the sign of the slow feedback loop dependent on a tunable balance between slow restorative and slow regenerative channels. Note that calcium currents have very low amplitude compared to the slow restorative potassium currents, so their modulation capabilities are located near the resting potential, where potassium channels are barely active. This property ensures that the positive feedback brought by slow regenerative channels does not interfere during the action potential downstroke. 

The ultraslow feedback gain accounts for the net sum of ultraslow currents. We focus in the present paper on situations where the equivalent ultraslow feedback gain is strictly negative. In the STG model, it corresponds to a situation where calcium-activated potassium channels have a dominant role in this timescale. The amplitude of this ultraslow feedback gain is nevertheless modulated by a balance between ultraslow restorative and ultraslow regenerative channels. 

To summarize, the many parallel feedback loops of any arbitrary conductance-based model aggregate in a fast positive feedback loop (autocatalysis), a slow positive-or-negative feedback loop (balance of neuronal excitability), and an ultraslow negative feedback loop (adaptation) in parallel with a static current. A rigorous mathematical analysis of this three time-scale structure is presented in Franci et al., 2013b. It shows that all the possible quantitative behaviors are captured by studying local variations around the particular situation where the slow feedback gain vanishes. Mathematically, this occurs at a transcritical bifurcation (Drion et al., 2012; Franci et al., 2013a). The minimal model used for our simulations only studies linear variations away from this bifurcation, with a linear gain $\bar{g}_s$ accounting for the slow feedback loop and a linear gain $\bar{g}_{us}$ accounting for the ultraslow feedback loop. We argue that much of neuronal modulation and robustness can be understood from these two parameters.

\subsection*{Spiking robustness arises from ion channel aggregation}
The equivalent conductances $\bar{g}_{s}$ and $\bar{g}_{us}$ determine an aggregate current in the slow and ultraslow timescales, respectively, regardless of involved ion channel specificities. This observation suggests strong robustness with respect to parameter variations that leave the equivalent conductances unchanged (see e.g. Taylor et al., 2009; Zhao et al., 2012) and it is consistent with the large variability measured experimentally (see e.g. Liu et al., 1998; Golowasch et al., 1999; Schultz et al., 2006; Marder, 2011).

\begin{figure}[h!]
\centering
 \includegraphics[width=0.95\linewidth]{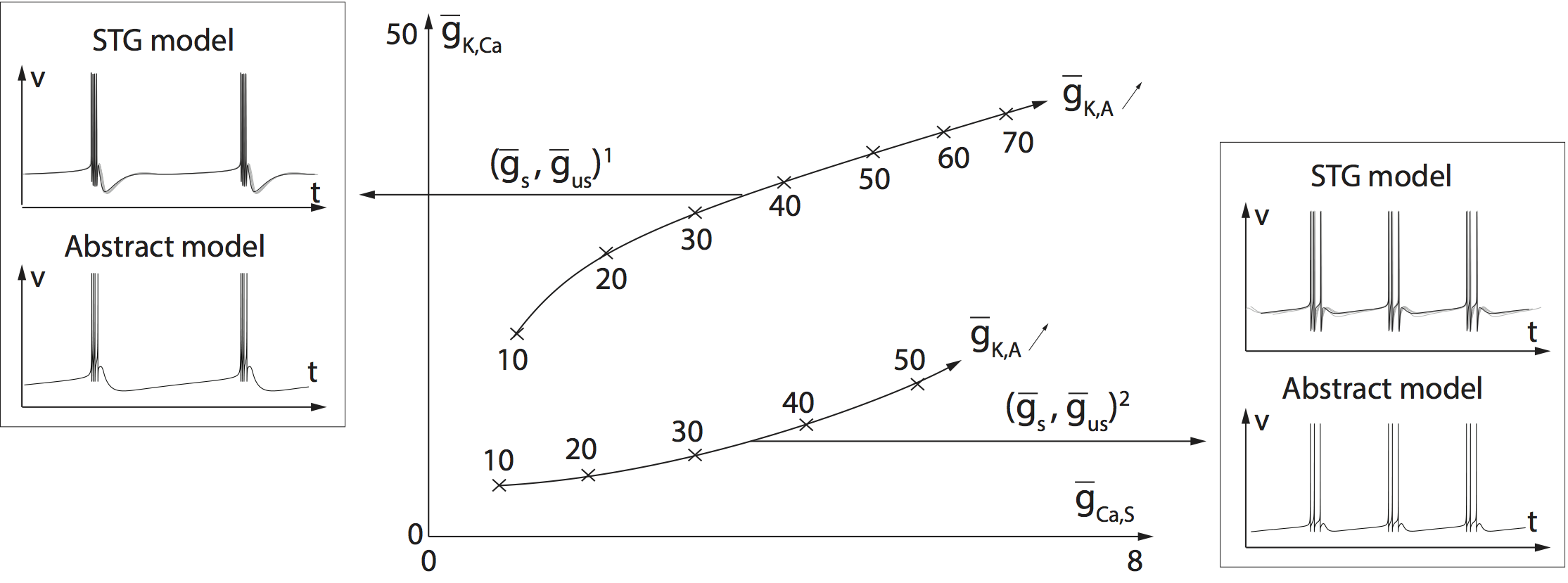}
\caption{\textbf{Different ion channel combinations that generate similar equivalent feedback gains lead to similar firing activities.} The central panel shows combinations of the three maximal conductances $(\bar{g}_{K,A},\bar{g}_{Ca,S},\bar{g}_{K,Ca})$ that lead to a same spiking behavior in the STG neuron model. The external panels show the superposition of the temporal traces of the STG model for the different values marked as crosses in the central panel (top) and the equivalent temporal trace generated by the abstract model for the two corresponding equivalent gain values (bottom). }\label{FIG:3}
\end{figure}

This robustness property is illustrated on the STG model in Figure \ref{FIG:3}, which shows that changes in the maximal conductance of one channel type (A-type potassium channels) can be balanced by a concomitant change in the maximal conductance of two others (slow calcium and calcium-activated potassium channels) to maintain specific firing activities. The temporal traces of each firing activity, which are superposed at the top of the left and right inserts of Figure \ref{FIG:3}, are barely discernible from each other, even though the maximal conductance of the potassium current $I_{K,A}$ is increased up to sevenfold. The abstract model predicts two different firing patterns for the two distinct couples of equivalent conductances, in good agreement with the STG neuron model.
 
The A-type potassium current is slow restorative (cf. Figure \ref{FIG:2}). Changes in its maximal conductance therefore affect the slow feedback gain $\bar{g}_s$, which results in a modulation of the firing activity of the cell. To maintain the initial firing activity, the variations of $\bar{g}_{K,A}$ have to be balanced by the variation of any other slow current maximal conductance. In Figure \ref{FIG:3}, the increase in the amplitude of the slow restorative current $I_{K,A}$ is compensated by an increase in the amplitude of the slow regenerative current $I_{Ca,S}$ to maintain the value of the slow feedback gain $\bar{g}_s$ (see Figure \ref{FIG:3}, center). Because activation kinetics of $I_{Ca,S}$ are slightly slower than the activation kinetics of $I_{K,A}$, the former current also mildly affect the ultraslow feedback gain $\bar{g}_{us}$. This effect is balanced by the modulation of the maximal conductance of the calcium-activated potassium ($\bar{g}_{K,Ca}$), which only contributes to the ultraslow feedback gain $\bar{g}_{us}$. 

One can therefore find many parameter combinations $(\bar{g}_{K,A},\bar{g}_{Ca,S},\bar{g}_{K,Ca})$ for which the values of $\bar{g}_s$ and $\bar{g}_{us}$, and thus the firing activity, are maintained. Two of such combinations leading to two different firing activities are illustrated in Figure \ref{FIG:3}, center. The two families of conductances correspond to two distinct couples of equivalent gains $(\bar{g}_{s},\bar{g}_{us})$ in the abstract model (left and right inserts of Figure \ref{FIG:3}, bottom).

This illustration shows that many different combination of ion channels can produce similar equivalent gain values, and therefore exhibit almost identical firing patterns. This simple property is at the basis of the robustness properties of neuronal signaling: a neuron is able to maintain a target firing activity despite changes in its environment and/or gene expression, as extensively observed experimentally (Namkung et al., 1998; Wickman et al., 1998; Brickley et al., 2001, Swensen and Bean, 2005). This property is crucial for intrinsic homeostasis, a central element in the preservation of nervous system functions (see e.g. Davis, 2006 and Turrigiano, 2011 for comprehensive reviews on the subject).

\subsection*{Modulation of the slow feedback gain: a route from slow spiking to bursting}
The slow feedback gain $\bar{g}_s$ results from a balance between the slow negative feedback of potassium currents and the slow positive feedback of calcium currents. If the density of calcium channels is low, then the net balance is largely negative and the spike generation is memoryless: the only role of slow currents is to restore the resting potential. This is the classical neuronal excitability of the HH model and most reduced spiking models. 

\begin{figure}[h!]
\centering
 \includegraphics[width=1\linewidth]{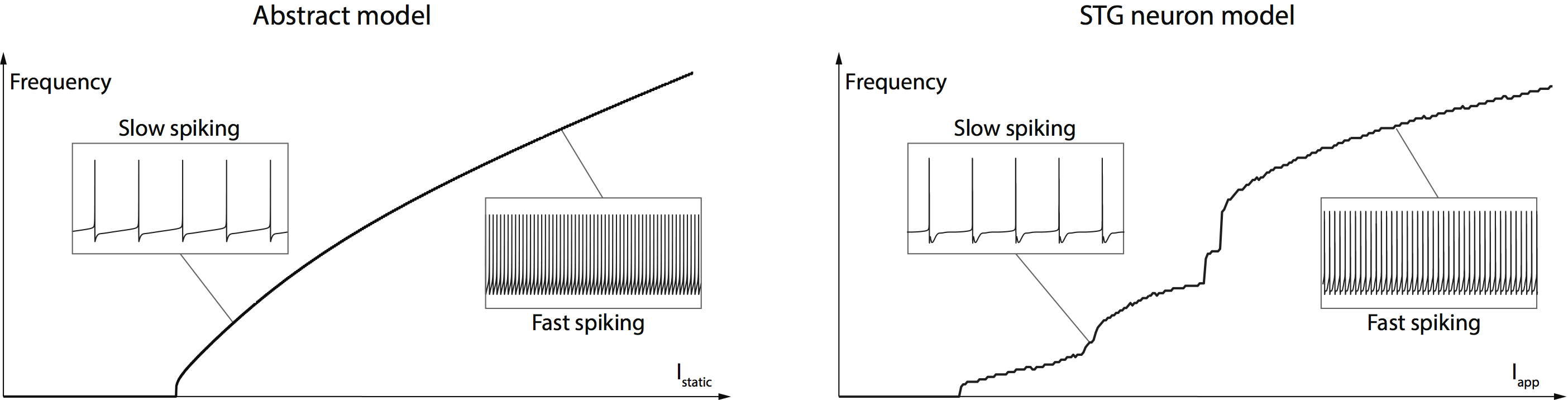}
\caption{\textbf{I/f curve of the abstract model (left) and the STG neuron model (right).}. Both models spike endogenously beyond a critical level of depolarization, with a frequency that monotonically increases with the static current.} \label{FIG:4}
\end{figure}

Figure \ref{FIG:4}, left illustrates this property in the abstract model. It shows that a three timescale model with a strictly negative slow feedback loop is qualitatively equivalent to a (type I) two timescale model: it can solely switch from quiescence to tonic spiking through variations of the static current $I_{static}$, and spiking frequency is dependent upon the amplitude of this static current. This property is also observed in the STG neuron model when the balance between slow restorative and slow regenerative ion channels is strictly negative (Figure \ref{FIG:4}, right). 

\begin{figure}[h!]
\centering
 \includegraphics[width=1\linewidth]{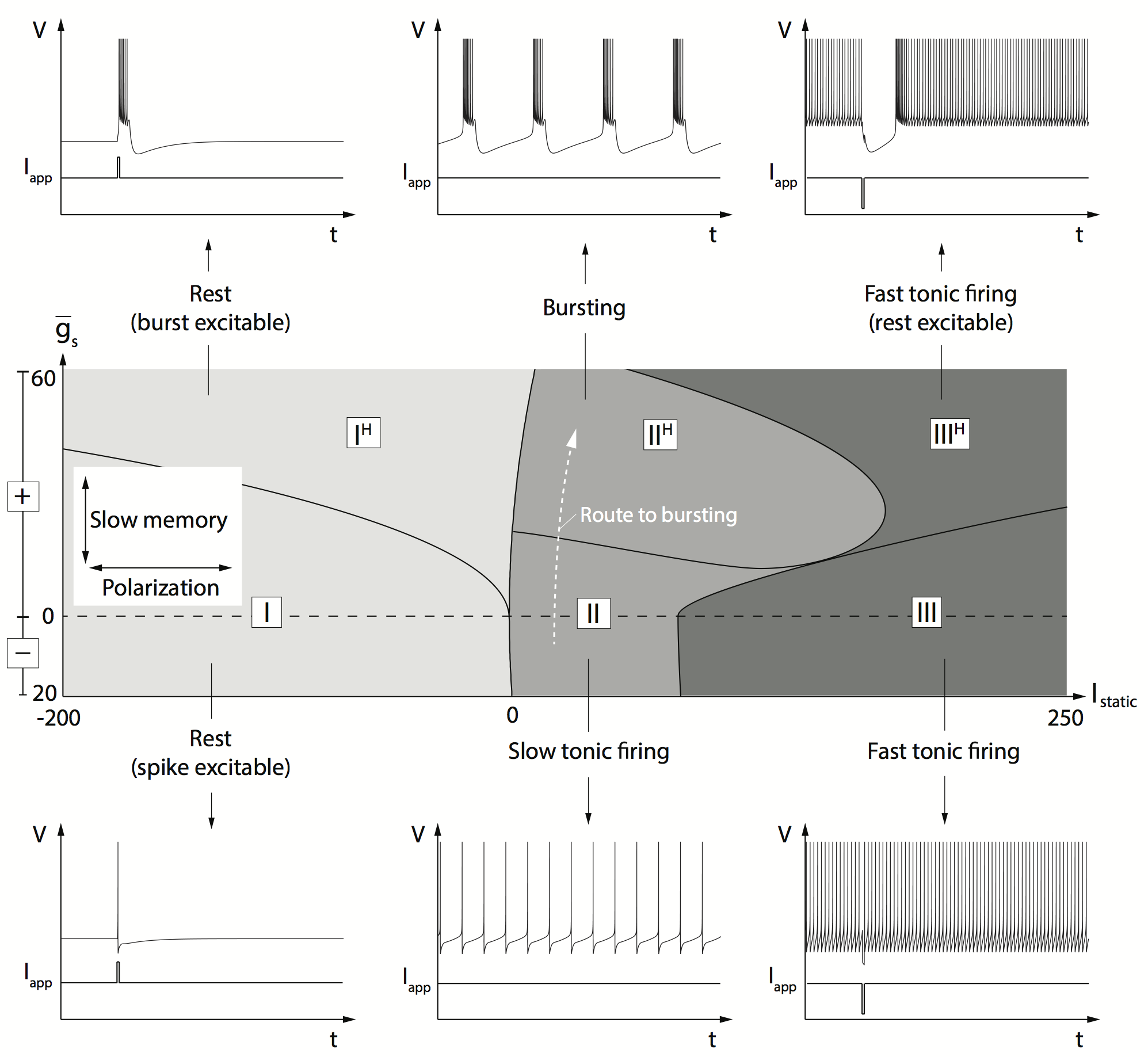}
\caption{\textbf{Modulation of the slow feedback gain $\bar{g}_s$ provides a route from slow tonic spiking to bursting.} The central two-parameter chart separates the firing activity in three different regions: quiescence (I), slow firing (II) and fast firing (III) (the sign of the slow feedback loop is given at the right hand-side of the panel). Each region shows two variants differing in their memory properties. The six external panels show illustrations of the six different firing patterns (all parameter values are available in the Methods section). $I_{static}$ is responsible for the switch between the three regions, whereas $\bar{g}_{s}$ mostly tunes the memory properties of neuronal spiking.} \label{FIG:5}
\end{figure}

When the density of calcium channels is high, it can revert the sign of the slow feedback loop and make it positive, but only locally around the resting potential. This is because the amplitude of the calcium currents is a tiny fraction of the amplitude of potassium currents, which means that calcium currents can revert the balance only at low potential where potassium channels are barely activated. A net feedback that is positive at low membrane potentials and negative at high membrane potential is a robust mechanism for rest and spike bistability (Franci et al., 2012; Franci et al., 2013a,b). Because of the slow positive feedback, spike generation acquires memory in the slow time-scale. The resulting neuronal excitability is illustrated in Figure \ref{FIG:5}, top: the memory in the slow time-scale is visible in the conversion of spikes into bursts and in the duration of the transient response to depolarizing or hyperpolarizing current pulses.

The six distinct regions in Figure \ref{FIG:5} cover six firing patterns of importance in electrophysiological recordings. Modulation along the horizontal axis is a well known feature of the polarization level of the resting potential. Modulation along the vertical axis is an important route from slow tonic firing to bursting. It is rooted in a switch from restorative excitability to regenerative excitability found in many neuronal models (Franci et al., 2013a). 

\begin{figure}[h!]
\centering
 \includegraphics[width=1\linewidth]{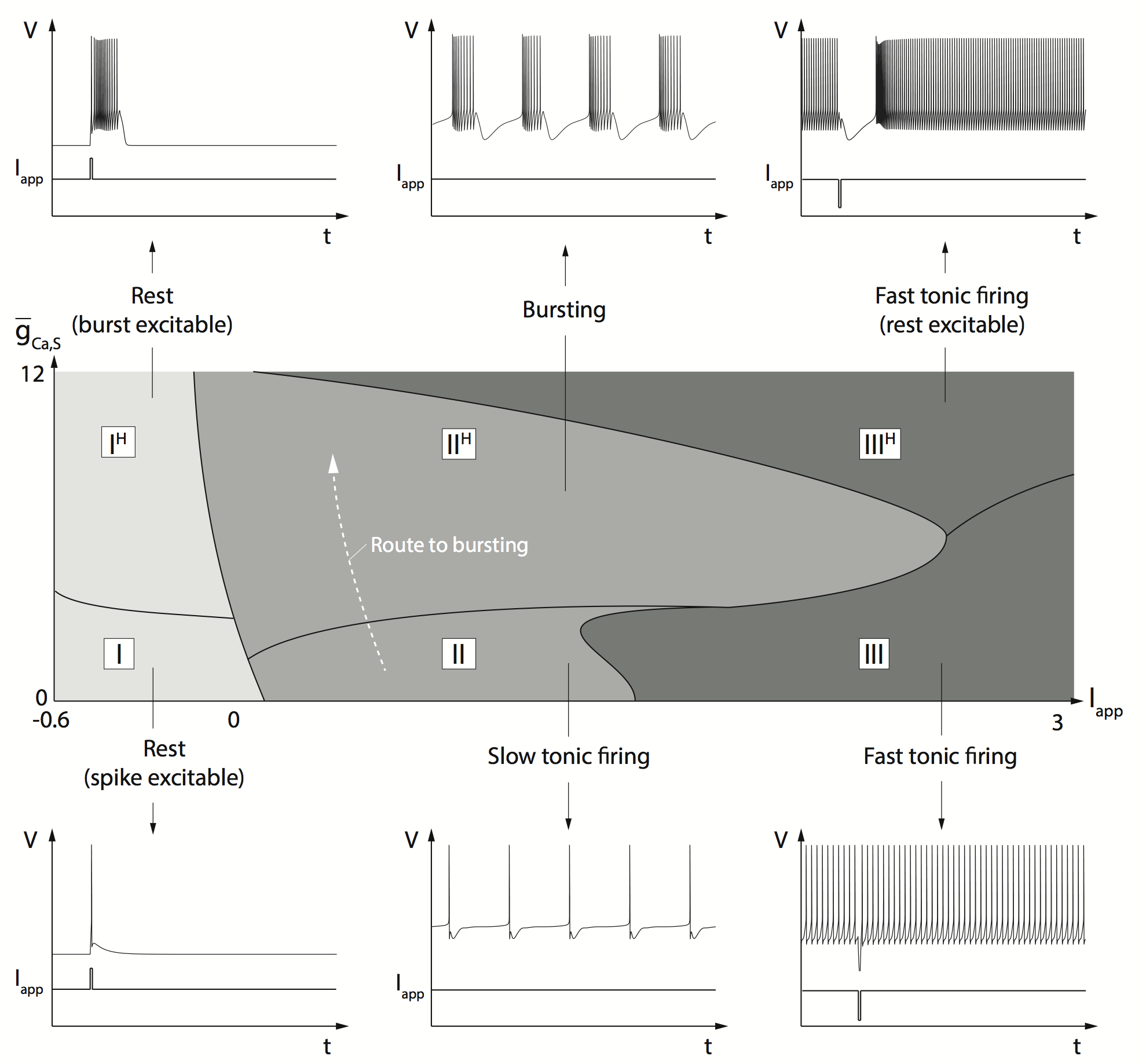}
\caption{\textbf{Variations of $\bar{g}_{Ca,S}$ and $I_{app}$ in the STG neuron model shows a classification of firing patterns qualitatively similar to the one predicted by the abstract model.} The central two-parameter chart separates the firing activity in three different regions: quiescence (I), slow firing (II) and fast firing (III). Each region shows two variants differing in their memory properties. The six external panels show illustrations of the six different firing patterns (all parameter values are available in the Methods section). In this example, $I_{app}$ is responsible for the switch between the three regions, whereas $\bar{g}_{Ca,S}$ mostly tunes the memory properties of neuronal spiking.}\label{FIG:6}
\end{figure}

In Figure \ref{FIG:6}, we reproduce the abstract prediction of Figure \ref{FIG:5} in the STG neuron model. The static current (horizontal axis) is modulated by changing the applied current $I_{app}$ while the slow feedback gain is modulated by varying the calcium maximal conductance $\bar{g}_{Ca,S}$. At the qualitative level, the two figures have striking similarities. Replacing $\bar{g}_{Ca,S}$ by the maximal conductance of any other slow current provides the same result (not shown).

\subsection*{Modulation of bursting: hysteresis and adaptation}
\begin{figure}[h!]
\centering
 \includegraphics[width=1\linewidth]{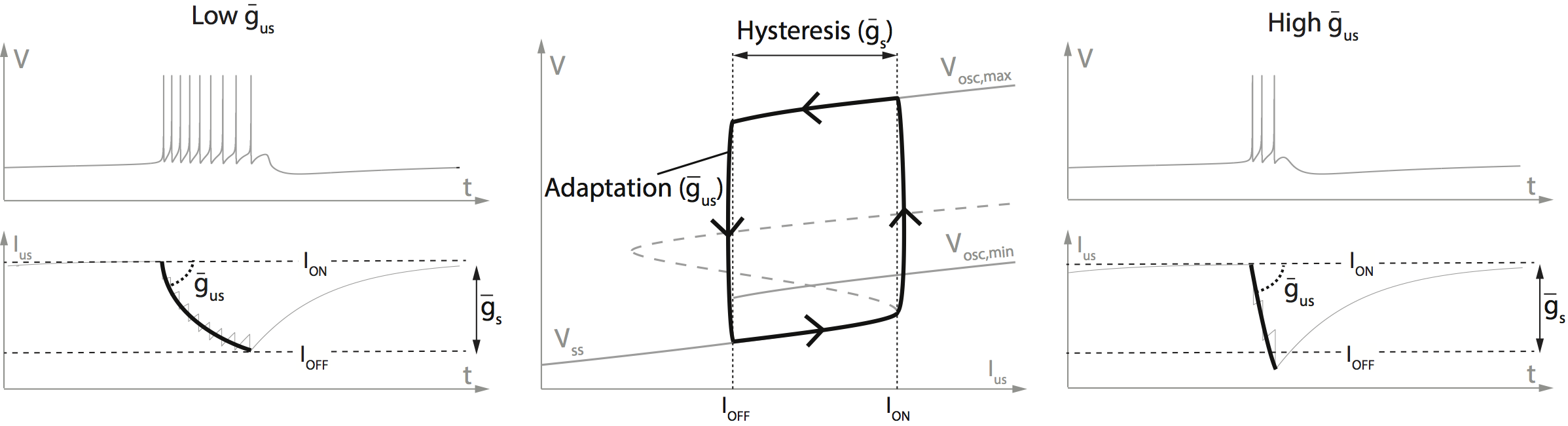}
\caption{\textbf{ Modulation of bursting by the slow and ultraslow equivalent conductances.} Bursting is schematized by a hysteretic loop in an I/V diagram (center). The silent period corresponds to a slow depolarizing drift along the lower branch of the hysteresis. The burst occurs on the upper branch, each spike causing an increment of hyperpolarizing current. The slow feedback gain determines the hysteretic range and the intraburst frequency. The ultraslow feedback gain determines the speed at which the hysteretic loop is traversed. Higher values of the ultraslow equivalent conductance translate into a shorter burst with less spikes (compare left and right figures).}\label{FIG:7}
\end{figure}

The route to bursting described in the previous section is highly plastic. It can be modulated in many ways, leading to a variety of bursting types in physiological recordings. In our abstract model, this modulation is aggregated in the slow and ultraslow feedback gains. Further insight in their respective contribution for the modulation of bursting comes from the representation of bursting as a hysteretic path in the I/V diagram of Figure \ref{FIG:7}. The silent phase of bursting corresponds to a slow drift along the low branch of the hysteresis, driven by an ultraslow depolarizing current. The spiking phase of bursting corresponds to the reverse drift along the upper branch of the hysteresis, driven by the hyperpolarizing increment of current resulting from each spike. Two parameters primarily determine the bursting shape in this diagram: the hysteretic range of the current modulation and the speed at which the hysteresis is looped under the drive of ultraslow currents. A larger hysteresis increases the contrast between silent and spiking periods of the burst, but the interburst frequency of spiking is highly dependent of the adaptation gain on the hysteretic loop. 

In our aggregate model, the slow feedback gain $\bar{g}_s$ of the regenerative slow feedback influences both the hysteretic range and the spiking frequency: the hysteretic range vanishes at the balance (i.e. there is no hysteresis when the slow feedback is restorative) and increases with the gain of the slow regenerative feedback. Concomitantly, a larger slow equivalent gain decreases the interspike interval and therefore contributes to a higher intraburst frequency.

In contrast, the ultraslow feedback gain $\bar{g}_{us}$ primarily influences the adaptation gain of the hysteretic loop. For a fixed value of the slow gain (that is, fixing the intraburst frequency and the hysteretic range), an increasing ultraslow gain will increase the adaptation gain of the hysteretic loop because each spike will produce a larger increment of the ultraslow inhibitory current. The contrast between a low and high ultraslow feedback gains for a given value of the slow equivalent conductance is illustrated in Figure \ref{FIG:7}.

\begin{figure}[h!]
\centering
 \includegraphics[width=1\linewidth]{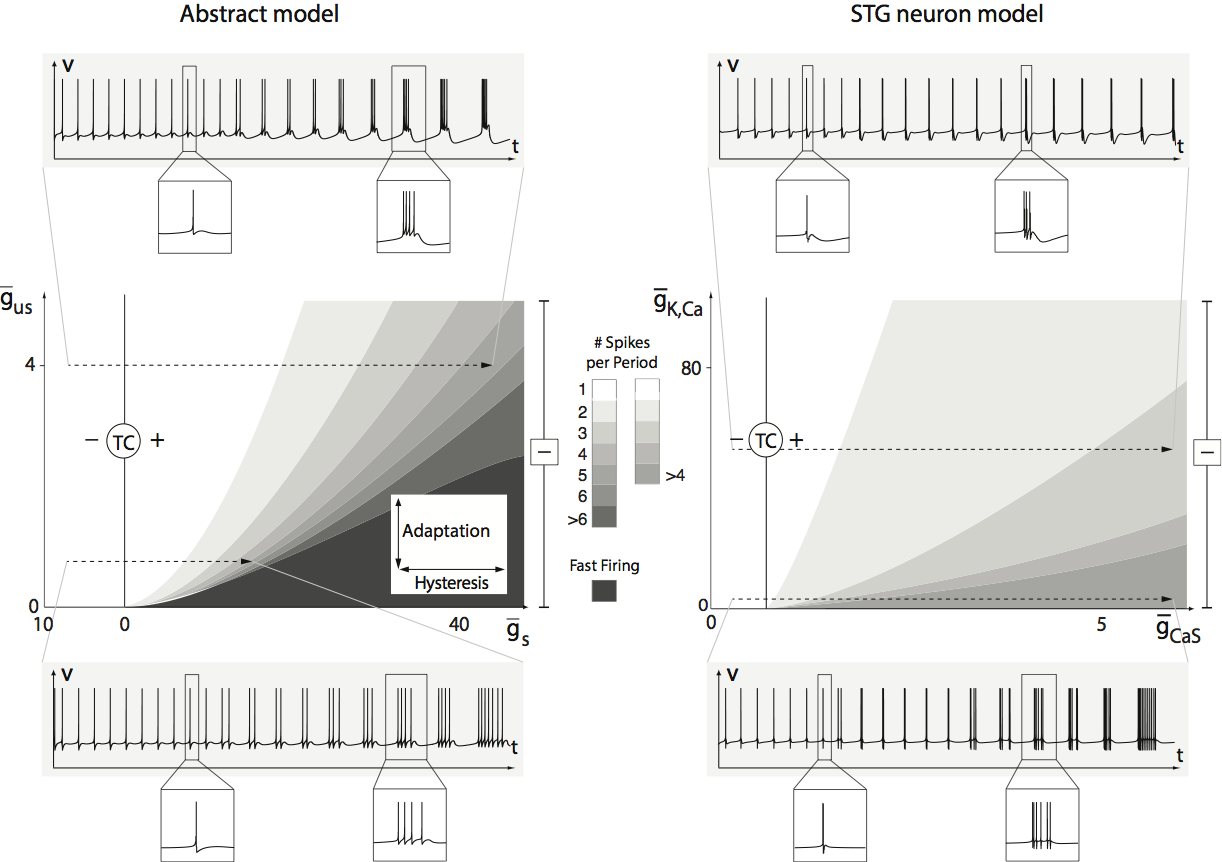}
\caption{\textbf{Routes to bursting in the abstract (left) and STG neuron model (right).} The central panels show the behavior of the abstract model (left) and the STG model (right) for different values of the slow and ultra-slow equivalent feedback gains (modulated through $\bar{g}_{Ca,S}$ and $\bar{g}_{K,Ca}$ in the STG model, respectively). The inserts show two routes to bursting for different values of $\bar{g}_{us}$.}\label{FIG:8}
\end{figure}

We illustrate in Figure \ref{FIG:8} the diversity of routes to bursting that can be obtained by modulating the slow ($\bar{g}_{s}$, horizontal axis) and ultraslow ($\bar{g}_{us}$, vertical axis) equivalent conductances in both the abstract model (left) and the STG neuron model (right). Each horizontal path in this diagram provides a route from slow tonic firing to bursting. The mathematical switch is through the transcritical bifurcation at $\bar{g}_{s} = 0$. The figure shows that for different values of the ultraslow gain $\bar{g}_{us}$, increasing the slow (regenerative) gain $\bar{g}_{s}$ leads to qualitatively different routes to bursting. It is interesting to note that slow tonic activities are barely differentiable from one configuration to the other, which highlights that two identical slow tonic spiking behaviors can hide two highly different bursters. This observation might have relevance in experimental electrophysiology.

\begin{figure}[h!]
\centering
 \includegraphics[width=1\linewidth]{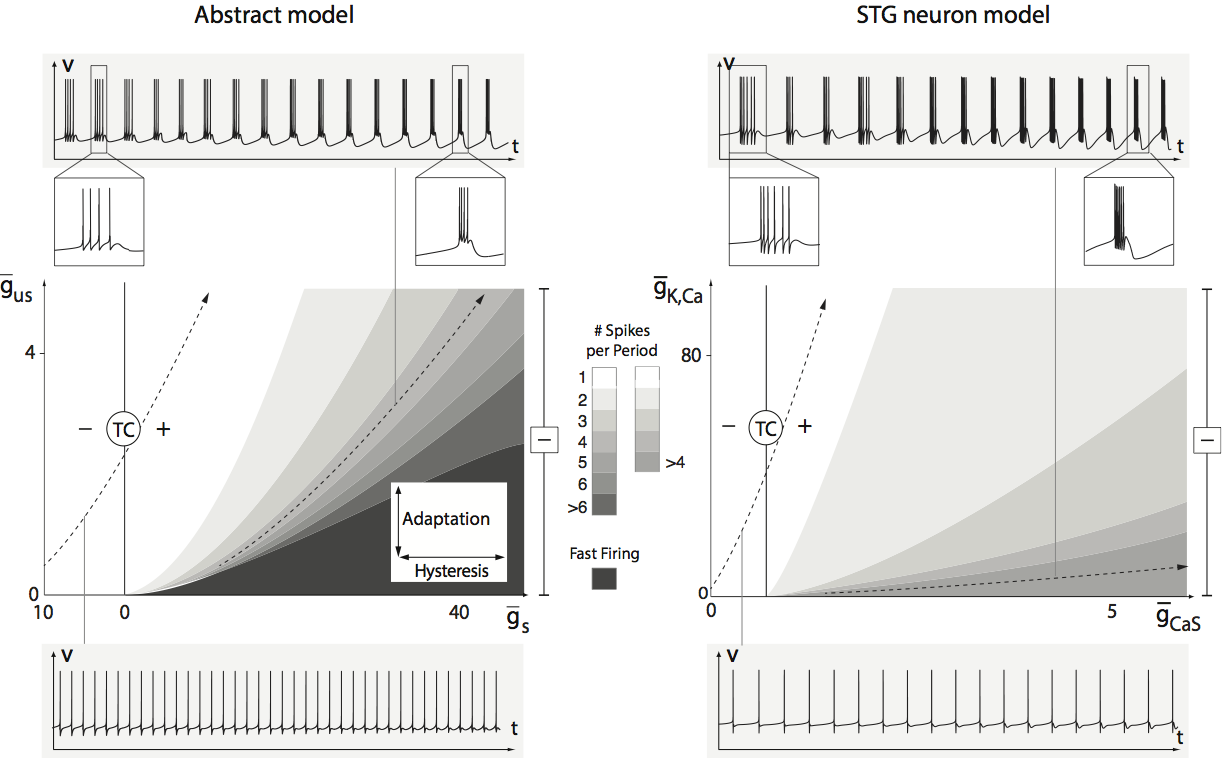}
\caption{\textbf{Modulation of firing patterns in the abstract (left) and STG neuron model (right).} The central panels show the behavior of the abstract model (left) and the STG model (right) for different values of the slow and ultra-slow equivalent feedback gains (modulated through $\bar{g}_{Ca,S}$ and $\bar{g}_{K,Ca}$ in the STG model, respectively). The inserts show modulation pathways for bursting (top) and slow tonic spiking (bottom). Although slow tonic spiking is very robust against variations in the values of the slow and ultraslow feedback gains, bursting can be strongly modulated.}\label{FIG:9}
\end{figure}

The existence of qualitatively different routes to bursting comes from the fact that the modulation of the two gains affects the burst wave shape (Figure \ref{FIG:9}). Low values of the regenerative gain $\bar{g}_s$ and the restorative gain $\bar{g}_{us}$ lead to non-plateau bursts with a low ratio between the intraburst and inter burst frequencies (Figure \ref{FIG:9}, left of upper panels). Such a wave form is reminiscent of the parabolic bursting recorded in Aplysia's R15 neurons in control conditions, see e.g. Rinzel and Lee, 1987; Levitan and Levitan, 1988. In contrast, high values of $\bar{g}_s$ and $\bar{g}_{us}$ lead to square-wave bursts with a high ratio between intraburst and interburst frequencies (Figure \ref{FIG:9}, right of upper panels). Such traces are reminiscent of electrophysiological recordings in many mammalian neurons, such as thalamocortical cells (McCormick et al., 1992; McCormick et al., 1997; Sherman et al., 2001) and subthalamic nucleus neurons (Beurrier et al., 1999; Hallworth et al., 2003), for instance. Interestingly, any neuron that is capable of bursting is also capable to modulate its bursting shape through the simple modulation of ion channel maximal conductances. This has for instance been observed in Aplysia R15 neurons under the action of serotonin (5-HT), as illustrated below. 

The significance of Figures \ref{FIG:8} and \ref{FIG:9} is not in the detailed simulation of a particular bursting trace but rather in the fact that a same modulation mechanism can account for many variations of electrophysiological signaling. Different ionic currents may lead to distinct bursting traces but through the same regulation mechanism of two equivalent feedback gains. This prediction is systematically verified in the STG neuron model. The maximal conductances of two particular currents are chosen as modulators of $\bar{g}_s$ and $\bar{g}_{us}$: the slow regenerative calcium current $I_{Ca,S}$ and the ultraslow restorative calcium-activated potassium current $I_{K,Ca}$. The modulation capabilities of these two parameters is in striking agreement with the predictions of the abstract model. The transcritical bifurcation of the abstract model is recovered in the full STG model at a critical value of $\bar{g}_{Ca,S}$. It delineates the same transition between slow tonic spiking ($\bar{g}_{Ca,S}<\bar{g}_{Ca,S}^{TC}$) and bursting ($\bar{g}_{Ca,S}>\bar{g}_{Ca,S}^{TC}$). Likewise, the modulation from ``parabolic-like'' bursting in Figure \ref{FIG:9} to ``square-wave'' bursting follows the prediction of the abstract model. Similar charts are obtained when replacing $\bar{g}_{Ca,S}$ by the maximal conductance of any other slow current (not shown).

\subsection*{Transient neuromodulation by T-type calcium channels}
T-type calcium channels probably figure amongst the most studied endogenous regulators of neuronal activity. We will use the parameter chart of Figure \ref{FIG:5} to revisit three experimental evidences of their neuromodulation properties: hyperpolarization-induced bursting (see e.g. McCormick and Bal, 1997; Beurrier et al., 1999; Hallworth et al., 2003), hyperpolarization-induced modulation of input/output properties (see e.g. Jahnsen and Llinas, 1984; Beurrier et al., 1999; Sherman et al., 2001) and rebound bursting (see e.g. Jahnsen and Llinas, 1984; Bevan et al., 2000; Astori et al., 2011). 

\begin{figure}[h!]
\centering
 \includegraphics[width=0.95\linewidth]{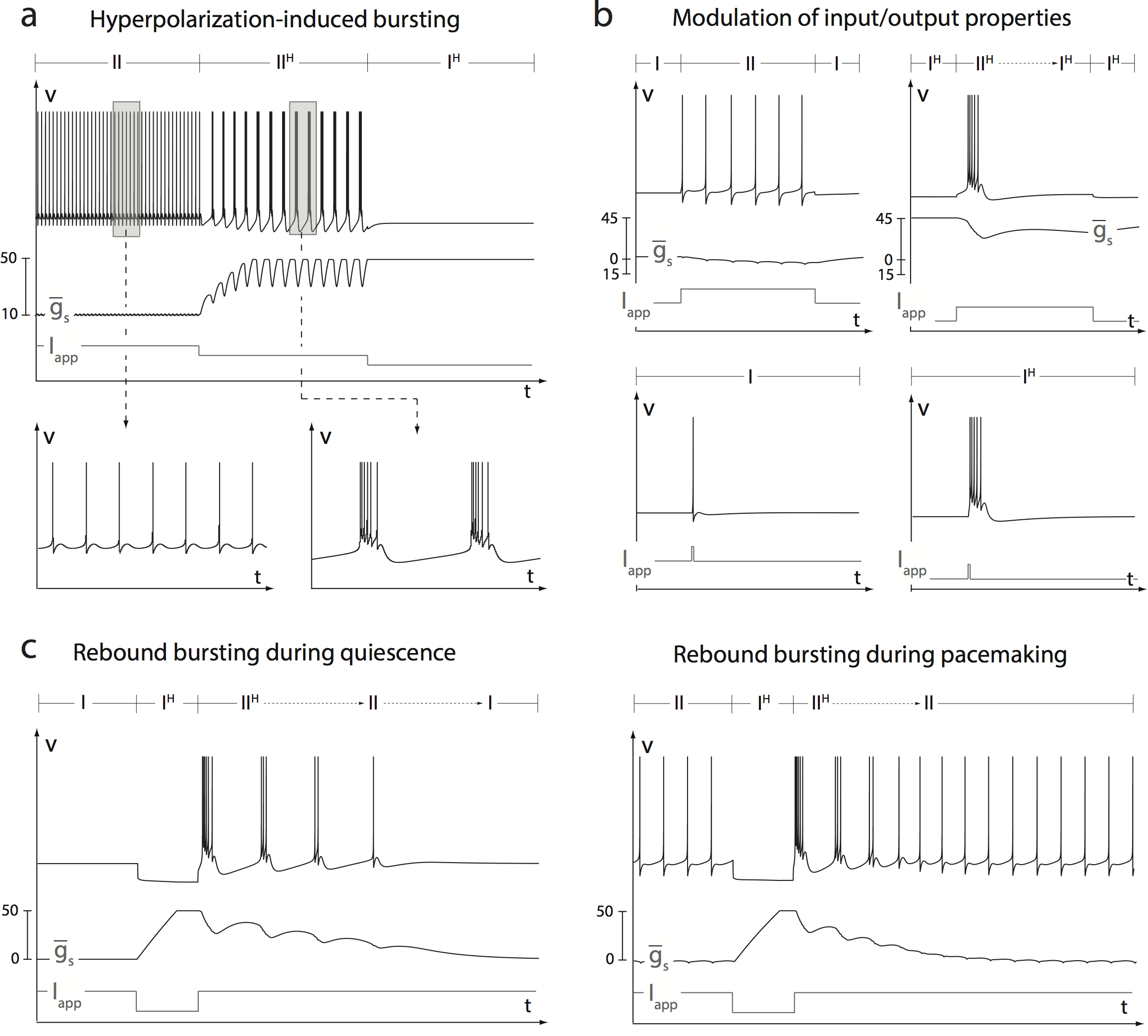}
\caption{\textbf{Transient neuromodulation by T-type calcium channels.} The figure shows the predictions of the abstract mode in different modulation scenarii. \textbf{a.} Responses of a spontaneously active neuron to steps of hyperpolarized current. Mild hyperpolarization induces a switch from slow tonic firing to bursting with an intraburst frequency that is much higher than the slow tonic firing rate. This switch is due to the hyperpolarization-induced increase in $g_{s}$, mimicking the deinactivation of T-type calcium channels. Further hyperpolarization silences the neuron, which remains burst excitable. \textbf{b.} Responses of a silent neuron to steps (top) and pulses (bottom) of excitatory current starting from different resting potentials. The initial resting potential determine neuron response to external stimuli, due to the voltage-dependence of $g_{s}$. \textbf{c.} Rebound activity after a transient hyperpolarization in a silent (left) and a spontaneously active neuron (right). The hyperpolarization induces an increase in $g_{s}$ via T-type calcium channel deinactivation, which is at the origin of the rebound bursting that occurs after the release of the inhibitory current. The regions of the parameter chart of Figure \ref{FIG:5} corresponding to the different firing patterns are plotted at the top of each voltage trace.}\label{FIG:10}
\end{figure}

Calcium channels are slowly regenerative and ultraslow restorative but the particularity of T-type calcium channels is that their inactivation occurs at low threshold. In other words, they come into play only when the neuron is sufficiently hyperpolarized. Hyperpolarizing a membrane that contains a large density of T-type calcium channels deinactivates them in the ultraslow time-scale, uncovering their role in the slow time-scale. We model this simple mechanism in our abstract model through voltage-dependent adaptation of the slow equivalent gain in the ultraslow time-scale, mimicking the low threshold inactivation of the channels (see methods). The resulting time traces are illustrated in Figure \ref{FIG:10}, in striking similarity with the experimental recordings:

\begin{itemize}
\item Figure \ref{FIG:10}a illustrates hyperpolarization-induced bursting (see e.g. Figure 4, Figure 7a and Figure 7b of Beurrier et al., 1999 for experimental recordings in STN cells and Figure 2 of McCormick and Bal, 1997 for experimental recordings in thalamic relay cells). Two successive steps of hyperpolarizing current $I_{app}$ cause a switch from slow tonic firing to bursting and from bursting to rest, respectively. In Figure \ref{FIG:5}, this corresponds to a path $II-II^H-I^H$, which combines a decay of $\bar{g}_{static}$ (due to the hyperpolarizing current) and an increase of $\bar{g}_{s}$ (due to the deinactivation of T-type calcium channels). The reader will notice that the path $II-II^H$ accompanying the first step of applied current is dominantly vertical in Figure \ref{FIG:5}. This is because the static component of the (depolarizing) calcium current compensates for the step of applied (hyperpolarizing) current in $I_{static}$.
\item Figure \ref{FIG:10}b illustrates how the level of polarization affects the input/output response of the neuron (see e.g. Figure 2 of Jahnsen and Llinas, 1984 and Figure 7a of Beurrier et al., 1999). Without hyperpolarization, the excitability of the neuron is restorative (Figure \ref{FIG:10}b, left): spike excitability for a pulse of applied current and slow tonic firing for a step of applied current. With hyperpolarization, the excitability of the neuron is regenerative (Figure \ref{FIG:10}b, right): burst excitability for a pulse of applied current and bursting for a step of applied current. Note that the bursting activity is only transient because of the inactivation of T-type calcium channels in the ultraslow time-scale.
\item Figure \ref{FIG:10}c illustrates the phenomenon of rebound bursting (see e.g. Figure 3 of McCormick and Bal, 1997 and Figure 2 of Bevan et al., 2000): a transient hyperpolarizing current causes transient burst activity in a quiescent (left) or slow tonic firing (right) neuron. The corresponding path in Figure \ref{FIG:5} is a loop $I, I^H, II^H, II, I$ (left) or $II, I, I^H, II^H, II$ (right). The succession of events is: $I_{static} \searrow$ (hyperpolarization), $g_{s} \nearrow$ (deinactivation), $I_{static} \nearrow$ (depolarization), $g_{s} \searrow$ (inactivation). 
\end{itemize}

The three modulations reproduced above all reflect a single and same modulation mechanism achieved by a single type of ion channels. They correspond to distinct paths in the abstract parameter chart of Figure \ref{FIG:5}. One path in this abstract chart reflects one scenario of modulation of the static current and the slow equivalent gain but each abstract scenario can be realized with a variety of different ion channels in a variety of different neurons.

\subsection*{Long-lasting neuromodulation by metabotropic receptors}
\begin{figure}[h!]
\centering
 \includegraphics[width=0.95\linewidth]{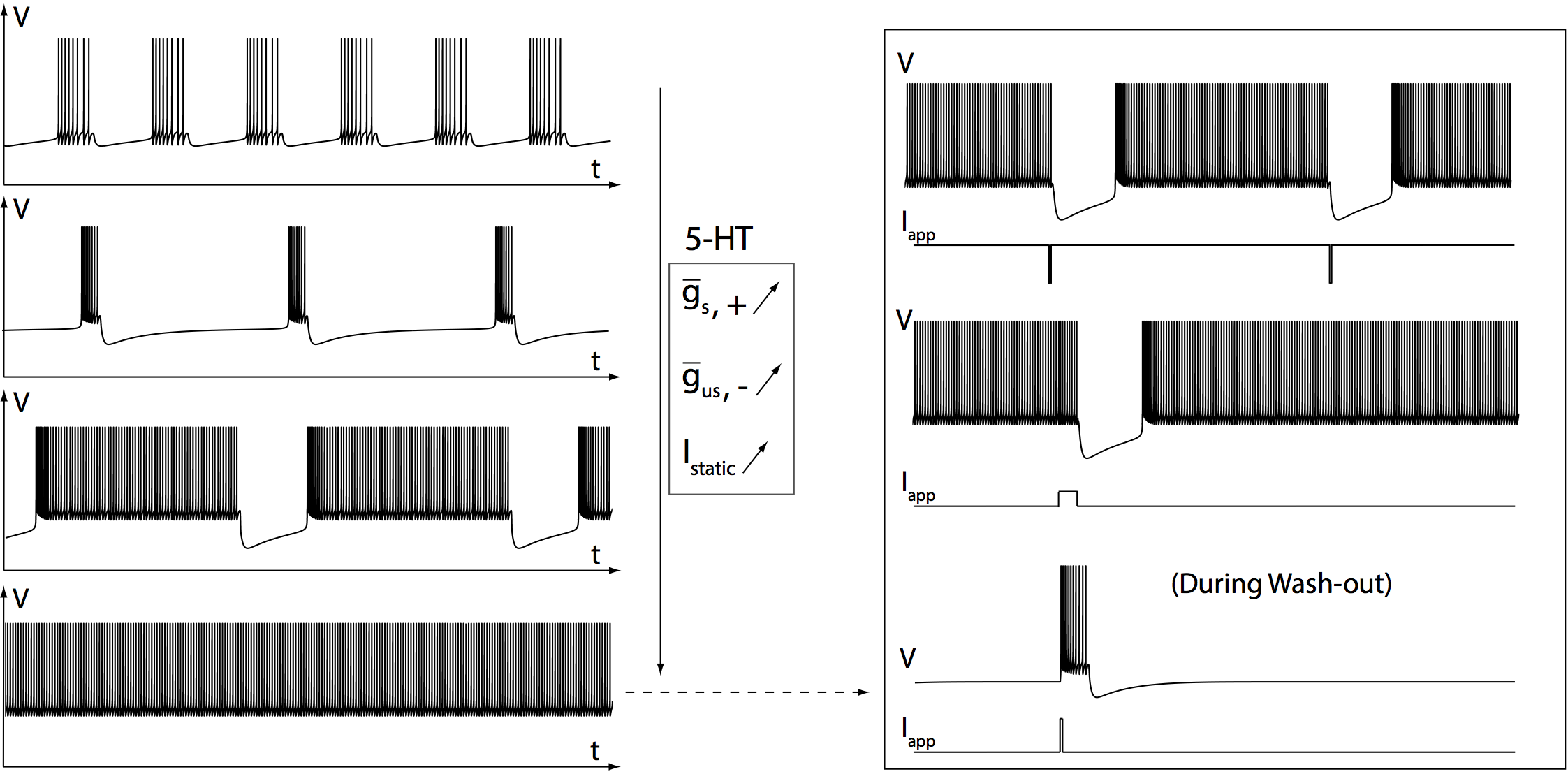}
\caption{\textbf{Illustration of Aplysia's R15 neuromodulation by serotonin (5-HT) in the abstract model.} \textbf{Left}, firing activity of the abstract model for increased concentrations of 5-HT (from top to bottom). Application of a moderate concentration of 5-HT induces a switch from parabolic-like to square-wave like bursting. Further increases in 5-HT concentration eventually result in fast tonic firing. \textbf{Right}, excitability properties of the neuron under high concentration of 5-HT (top, middle) and during wash-out (bottom). The neuron is strongly rest-excitable at high concentration of 5-HT, and becomes silent (burst excitable) during wash-out. These properties arise from an increase in $\bar{g}_{s}$ ($R_{s}=1$) and in $\bar{g}_{us}$ ($R_{us}=-1$) resulting from the effect of 5-HT on Aplysia's R15 ion channel densities (see Levitan and Levitan, 1988 for comparison and further experimental details).}\label{FIG:11}
\end{figure}

Metabotropic receptors are a well documented source of long-lasting neuromodulation. Located at the plasma membrane and sensitive to specific neurotransmitters, they modulate ion channel density and activation state, among others. We revisit in our abstract parameter chart an historical experimental evidence of the modulation of neuronal activity in Aplysia's R15 neuron by serotonin (5-HT). At the molecular level, 5-HT has been shown to increase the density of ultra-slowly activating potassium channels (Drummond et al., 1980; Benson and Levitan, 1983) and slowly activating calcium channels (Levitan and Levitan, 1988). We summarize this modulation in our abstract model by an increase in the static current $I_{static}$ (calcium channels are not fully deactivated at rest), an increase of the slow gain $\bar{g}_{s}, +$ (positive feedback loop) (calcium channels are slowly regenerative), and an increase of the ultraslow gain $\bar{g}_{us}$ (negative feedback loop, i.e. $R_{us}=1$) (potassium channels are ultraslow restorative). The resulting time traces in Figure \ref{FIG:11}, left (to be compared e.g. to the experimental trace in Figure 1 of Levitan and Levitan, 1988) are easily interpreted in the abstract parameter charts of Figure \ref{FIG:5} and Figure \ref{FIG:9}. The bursting pattern evolves from ``parabolic-like'' to ``square-wave'' as $\bar{g}_{s}$ and $\bar{g}_{us}$ increase. Quiescent periods of the bursts then decrease to eventually vanish when the static depolarization becomes sufficient. 

The same abstract model accounts for the particular excitability properties of the final fast tonic firing: short pulses of either hyperpolarizing or depolarizing current evoke long lasting periods of quiescence (Figure \ref{FIG:11}, right, to compare with the experimental trace in Figure 9c,e of Levitan and Levitan, 1988) and, likewise, bursts during wash-out induced hyperpolarized state (Figure \ref{FIG:11}, right, to compare with the experimental trace in Figure 9d of Levitan and Levitan, 1988). Those features are indeed signature of regions III$^H$ and I$^H$, respectively, and marly illustrate the regenerative nature of excitability at high concentration of serotonin. These signatures are in opposition with the effect of dopamine (DA), which has been shown to hyperpolarize the neuron (Gospe and Wilson, 1980) but does not induce burst excitability (Lotshaw and Levitan, 1988). This difference is due to the fact that Aplysia's R15 neuron reaches the restorative region I instead of the regenerative region I$^H$ under DA application, this neurotransmitter reducing the density of slow regenerative calcium channels (Lotshaw and Levitan, 1988).

\section*{Discussion}

\subsection*{Equivalent conductances determine firing activity}
Our paper advocates that the early methodology of Hodgkin and Huxley to dissect the mechanism of action potential generation extends with a comparable predictive power to dissect the mechanism of slow firing patterns over the course of many spikes. Specifically, we focus here on the feedback aggregated by many ion channels in the slow and ultraslow time-scales, determined by the intra and interburst frequencies, respectively.

\subsection*{Slow regenerativity acts as short term memory}
A highlight of our analysis is that the slow feedback gain can be either regenerative (positive sign) or restorative (negative sign) and that a regenerative slow feedback gain provides the neuron with short term memory, a hallmark of bursting. A mere modulation of the slow feedback gain, illustrated for instance in the present paper by the transient deinactivation of T-type calcium channels, provides a physiological route from slow tonic firing to bursting. To the best of the authors knowledge, this important signaling mechanism has not been described in previous mathematical models of neuronal bursting.

\subsection*{Bursting is a modulated signal}
The ``route to bursting'' mechanism described in the present study is not associated to one particular bursting type, but accounts for many of those previously observed experimentally. In contrast to the classification of bursters in different types of mathematical models, we view the mathematical modeling of bursting as one same geometric attractor modulated by few parameters (Franci et al., 2013b). In particular, a low ultraslow feedback gain results in a low ratio between the intraburst and inter burst frequencies, which is often described as parabolic bursting. On the other hand, a high ultraslow feedback gain results in square wave bursting with a high ratio between the intraburst and inter burst frequencies. Note that if the intraburst firing frequency is high enough to accumulate sodium channel inactivation, the neuron exhibits triangular bursting through the incremental decrease of the fast feedback gain during the bursts (not illustrated in the present paper). As a consequence, in addition to be able to switch from slow tonic firing to bursting, any bursting neuron is able to continuously evolve between different bursting types through the modulation of the slow and ultraslow feedback gains. One physiological instance of this modulation is through changes in ion channel densities, as it is observed in Aplysia's R15 neurons under the application of 5-HT.

\subsection*{Parameter redundance and intrinsic homeostasis}
The simple modulation mechanism described in this paper is of potential relevance to study intrinsic homeostasis. It indeed shows that neuron firing pattern is primarily determined by the values of the equivalent gains in each timescale, regardless of the combination of ion channels that generate them. Neurons can therefore realize one given firing activity through different combinations of channel densities, and the loss of one ion channel type can potentially be compensated for by other channels acting on the same timescale. Such adaptation properties have been repeatedly observed in experiments over the years (see e.g. Davis, 2006 and Turrigiano, 2011).

\subsection*{A physiologically rooted simple abstraction gathers a variety of physiological regulation mechanisms}
Several experimental observations are revisited in the paper. They span several time scales from spikes (ms) to intracellular processes (hours), and involve a variety of physiological mechanisms. They are interpreted as modulatory paths in the same two-parameter charts and are all simulated with the same low-dimensional computational model. The remarkable match between the abstract model predictions and the experimental observations does not owe to parameter tuning. Instead, it is rooted in the direct physiological interpretability of the proposed abstract parameters. Such an approach will hopefully contribute to a better understanding of system properties of neurons, including modulation, robustness and homeostasis of neuronal signaling.

\section{Materials and Methods}
\subsection{Abstract neuron model}
Our mathematical predictions use an abstract model of neuronal spiking based on the transcritical normal form (Drion et al., 2012)
\begin{eqnarray}\label{EQ: quantitative IZHw2 model}
\dot v=v^2+bvx_s-{x_s}^2\pm\mathbf{\bar{g}_s}x_s-\mathbf{\bar{g}_{us}}x_{us}+\mathbf{I_{static}}+I_{app}&\quad\quad\quad \mbox{if }\ v\geq v_{th},\ \mbox{then}\nonumber\\
\tau_{s}\dot x_s=a_sv-x_s&\quad\quad\quad v\leftarrow c,\ x_s\leftarrow d_s,\nonumber\\
\tau_{us}\dot x_{us}=a_{us}v-x_{us}&\quad\quad\quad x_{us}\leftarrow x_{us}+d_{us}\nonumber.
\end{eqnarray}
where $v$ merges the membrane potential and fast variables, $x_s$ merges all slow recovery variables and $x_{us}$ all ultra-slow adaptation variables. The only modulated parameters are the equivalent gains $\bar{g}_s$, $\bar{g}_{us}$ and the static current $I_{static}$ ($\bar{g}_s, \bar{g}_{us} \geq 0$). $I_{app}$ represents the applied current. $\bar{g}_s$ represents the slow equivalent gain, physiologically modulated by the balance between restorative and regenerative channels: the positive sign models the case where regenerative channels are dominant, the negative sign models the case where restorative channels are dominant, and $\bar{g}_s=0$ corresponds to an exact balance between restorative and regenerative channels. $\bar{g}_{us}$ is similar to $\bar{g}_s$ but in the timescale of ultra-slow ion channels. $I_{static}$ is the static current that determines the resting potential. These three parameters are physiologically regulated (Franci et al., 2013a), and are therefore varied over a wide range throughout the paper (particular values are given case by case below). All other parameters, such as reset values, are kept at a fixed value throughout the paper. The chosen values are the following: $b=-2$, $\tau_{s}=1$, $\tau_{us}=10$, $a_{s}=a_{us}=0.1$, $v_{th}=80$, $c=15$, $d_s=30$ and $d_{us}=20$.

To model the effect of T-type calcium channels in Figure \ref{FIG:10}, we model the kinetics of their ultraslow inactivation at low threshold ($g_s = \pm\bar{g}_s$), that is: 
\begin{eqnarray}\label{EQ: quantitative IZHw2 model}
\tau_{g_s}\dot{g}_{s}=a_{g_s}(v-v_{\frac{1}{2}})-g_{s} + g_{s,min} \nonumber
\end{eqnarray}
where $v_{\frac{1}{2}}$ is the inactivation threshold and $g_{s,min}$ the minimal value of $g_{s}$ reached when all virtual T-type calcium channels are inactivated. $g_s$ is also restricted to a maximal value $g_{s,max}$ through the saturation rule 
$$\mbox{if }\ g_{s} \geq g_{s,max},\ \mbox{then}\ g_{s} = g_{s,max}.$$
This results in a piecewise linear inactivation function centered at $v_{\frac{1}{2}}\geq0$ and whose minimal and maximal values are $g_{s,min}$ and $g_{s,max}$, respectively. The values of the fixed parameters are the following: $\tau_{g_s}=62.5$, $a_{g_s}=-13$, $g_{s,min}=-40$ and $g_{s,max}=50$. $v_{\frac{1}{2}}$ is set to 3 to model hyperpolarization-induced bursting and to 0 in any other case. Finally, we use the simple relation $I_{static}=\mid g_s\mid$ to model the effect of T-type calcium channel inactivation on the static current.

\subsection{STG neuron model}
Membrane currents are described in Goldman et al., 2001, whose model itself derives from Liu et al., 1998. The kinetics and voltage dependence of the conductances contributing to the currents are based on measurements of crustacean STG neurons Turrigiano et al., 1995. All parameters are similar to the ones given in Goldman et al., 2001, except for the calcium reversal potential which is fixed here to $-120$ mV. The model is composed of six different ionic currents: $I_{Na}$, $I_{Ca,T}$, $I_{Ca,S}$, $I_{K,A}$, $I_{K,d}$ and $I_{K,Ca}$. Following our classification, $I_{Na}$ is fast regenerative, $I_{Ca,T}$ and $I_{Ca,S}$ are both slow regenerative, $I_{K,A}$ is slow restorative, $I_{K,d}$ is slow restorative and $I_{K,Ca}$ is ultraslow restorative.

\subsection{Construction of the $I_{static} - \bar{g}_s$ parameter chart (Figure \ref{FIG:5})}
Figure \ref{FIG:5} is obtained using $\bar{g}_{us} = -3$. $\bar{g}_{s}$ is varied from $-20$ to $60$ by steps of $1$. For each value of $\bar{g}_{s}$, $I_{static}$ is varied from $-200$ to $250$ by steps of $1$. 

Threshold values are defined as follows: the threshold separating zones $I$ and $I^H$ corresponds to a saddle-node (SN) bifurcation of the $v-x_s$ subsystem. The threshold separating zones $I$ and $II$ corresponds to a saddle-node on invariant circle (SNIC) bifurcation of the $v-x_s$ subsystem. The threshold separating zones $I^H$ and $II/II^H$ corresponds to a saddle-homoclinic (SH) bifurcation of the $v-x_s$ subsystem. The threshold separating zones $II$ and $II^H$ corresponds to an abrupt increase in the coefficient of variation (CV) of the interspike intervals as $I_{static}$ increases. The threshold separating zones $II^H$ and $III^H$ corresponds to an abrupt decrease in the CV of the interspike intervals as $I_{static}$ increases. In both cases, the value $CV = 0.1$ is arbitrary taken to define the threshold. The threshold separating zones $II$ and $III$ corresponds to the value at which afterdepolarization potentials (ADP's) vanishes during tonic spiking for $\bar{g}_s > 0$ and it is extrapolated for $\bar{g}_s < 0$. From a dynamical viewpoint, slow tonic spiking corresponds to a situation where the activation of the ultraslow variable $x_{us}$ induces the crossing of the SNIC ($\bar{g}_s < 0$) or SN ($\bar{g}_s > 0$) bifurcations, whereas during fast tonic spiking the system remains on the stable limit cycle without crossing any bifurcation. Finally, the threshold separating zones $III$ and $III^H$ is computed by hand, zone $III^H$ corresponding to the values where a transient hyperpolarization is followed by a biphasic behavior, that is first hyperpolarization then depolarization toward fast spiking.

The values generating the time-courses shown in the inserts are the following: $(\bar{g}_{s}, I_{static}, I_{app})$ = $(-10, -10, 50)$ for spike excitability, $(-10, -30, /)$ for slow tonic spiking, $(-10, 400, -800)$ for fast tonic spiking, $(50, -10, 50)$ for burst excitability, $(50, 30, /)$ for bursting and $(50, 100, -200)$ for rest excitability. $I_{app}$ corresponds to the value of the impulse and is zero elsewhere.

\subsection{Construction of the $I_{app} - \bar{g}_{Ca,S}$ parameter chart (Figure \ref{FIG:6})}
Figure \ref{FIG:6} is obtained using the following maximal conductance values (in mS/cm$^2$): $\bar{g}_{Na} = 700$, $\bar{g}_{Ca,T} = 3$, $\bar{g}_{A} = 90$, $\bar{g}_{Kd} = 70$ and $\bar{g}_{K,Ca} = 20$. $\bar{g}_{Ca,S}$ is varied from $0$ to $5$ by steps of $0.5$ and from $5$ to $12$ by steps of $1$, all in mS/cm$^2$. For each value of $\bar{g}_{Ca,S}$, $I_{app}$ is varied by hand to compute the threshold values. 

All thresholds are computed by hand and are defined as follows: the threshold separating zones $I/I^H$ and $II/II^H$ corresponds to the values at which the model switch from quiescence to spiking. The threshold separating zones $I$ and $I^H$ corresponds to the values at which the model starts to generate bursts in response to applied depolarizing pulses. The threshold separating zones $II$ and $II^H$ corresponds to the values at which the model starts to exhibit doublets or bursts. The threshold separating zones $II$ and $III$ corresponds to the values at which afterdepolarization potentials (ADP's) vanishes during tonic spiking. The threshold separating zones $II^H$ and $III/III^H$ corresponds to the values at which the model switches from bursting to fast tonic spiking. Finally, the threshold separating zones $III$ and $III^H$ is defined as in the hybrid model.

The values generating the time-courses shown in the inserts are the following: $(\bar{g}_{Ca,S}, I_{app}, I_{app,2})$ = $(0, -0.3, 4)$ for spike excitability, $(1, 0.5, /)$ for slow tonic spiking, $(1, 1, -10)$ for fast tonic spiking, $(5, -0.3, 4)$ for burst excitability, $(7, 0.1, /)$ for bursting and $(12, 0.1, -10)$ for rest excitability. $I_{app,2}$ corresponds to the value of the impulse. $\bar{g}_{Ca,S}$ is in mS/cm$^2$ and $I_{app}/I_{app,2}$ are in $\mu$A/cm$^2$.

\section*{Acknowledgments}
This work is supported by grant 9.4560.03 from the F.R.S.-FNRS (VS) and by two grants from the Belgian Science Policy (IAP7/10 (VS) and IAP7/19 (RS)). The authors gratefully acknowledge several constructive discussions with Dr. Pierre-Alexandre Bliman (INRIA Rocquencourt) and Pr. Pierre Maquet (University of Li\`ege). The authors declare no competing financial interests.

\end{document}